# 3D Genetic Metamaterials for Scattering Maximization

Dmitry Dobrykh[1,*], Anna Mikhailovskaya[1], Konstantin Grotov[1], Dmytro Vovchuk[1], Vladyslav Tkach[2], Mykola Khobzei[2], Anton Kharchevskii[1], Aviel Glam[3], and Pavel Ginzburg[1]

[1] School of Electrical Engineering, Tel Aviv University, Tel Aviv 69978, Israel

[2] Department of Radio Engineering and Information Security, Yuriy Fedkovych Chernivtsi National University, Chernivtsi, 58000, Ukraine

[3] Rafael Advanced Defense Systems Ltd., Haifa, Israel

* dmitryd@mail.tau.ac.il

**Abstract** – The rapidly growing volume of drone air traffic demands improved radar surveillance systems and increased detection reliability in challenging conditions. The scattering cross-section, which characterizes a target's radar visibility, is a key element in detection schemes and thus becomes a primary objective in civilian applications. Here, we introduce a concept of genetically designed metamaterials, specifically engineered to enhance scattering for end-fire incidence scenarios. Multi-layer stacks of arrays, encompassing strongly coupled electric and magnetic resonators, demonstrated above 1 m$^2$ broadband scattering at 10 GHz, despite having an end-fire physical cross-section smaller than one squared wavelength. Those performances, crucial for effective civil radar air traffic monitoring, facilitate exploring highly scattering structures as labels for small airborne targets. This objective has been demonstrated with a set of outdoor experiments with the DJI Mini 2 drone. Lightweight, conformal add-ons with significantly high scattering cross-sections can serve as auxiliary tools to empower passive monitoring systems, thereby providing an additional layer of security in urban airspace.

# Introduction

The exploration of electromagnetic wave interactions with matter has been a focal point in both fundamental and applied research for a long time and, it keeps developing owing to emerging challenges and opportunities in fields such as telecommunications, medical imaging, and radar technology. The field of metamaterials emerged as a pathway to tailor the propagation of electromagnetic waves almost on demand, with the demonstration of a negative index being one of the first peculiar instances of this concept.[1–8] In the realm of scattering management, negative index metamaterials have been proposed and demonstrated to conceal objects from radar detection by effectively bending waves around an object, thus preventing interaction.[9–12] Admitting the complexity of electromagnetic designs, homogenization approaches for extracting effective material parameters were initially applied to identify the parameters of resonator arrays, typically encompassing both electric and magnetic resonators.[13–15] Only then were these metamaterials applied to an object subject to cloaking. However, the continuously growing computational capabilities, equipped with cutting-edge machine learning algorithms, allow for the exploration and optimization of more complex designs.[16–25] This approach enables addressing application needs upfront, bypassing the traditional reliance on bottom-up design principles. Hereinafter, our focus will be on the development of a volumetric metamaterial, to maximize electromagnetic scattering with a sufficient bandwidth to fit a range of emerging wireless applications.

The ever-increasing demand for wireless data traffic keeps challenging front-end hardware, thus making tailoring electromagnetic field propagation in complex conditions of heavy clutter the primary task. On the applied side, the focus will be on the forthcoming big issue of small unmanned aerial vehicle (UAV) monitoring in urban environments with radar technologies. Small drones are extremely challenging targets for radar detection and thus are demanding cooperation with monitoring authorities to ensure a high level of reliability.[26] While equipping UAVs with active transponders is already enforced by regulations, auxiliary passive labeling can significantly enhance monitoring reliability as it does not depend either on power-consuming electronics or the demand to maintain continuous wireless data traffic.[27] The design of electromagnetic structures for enhancing a small UAV radar scattering cross-section (RCS) is subject to many practical contractions, including aerodynamical compliance, weight limitations, and size conformity to name just a few. Considering the above-mentioned application, the control of scattering from objects under constraints remains a fundamental objective and ever-evolving challenge in applied electromagnetism.

Retroreflectors are typically the first choice in case a high RCS for a wide range of angles is required. The most widely used architectures, addressing these demands are corner reflectors, Luneburg lenses, and more advanced Van-Atta arrays.[28–34] However, based on geometrical optics approaches in the first case and periodic antenna arrays in the last, those structures are significantly larger than a wavelength and thus less likely to meet the above-discussed constraints. For example, a five-wavelength across moderate-size corner reflector will occupy more than 15 cm in each direction, given 10 GHz (X-band) radar system is in use. This observation raises a demand to design subwavelengths (or close to that) structures, which will still possess significant RCS. From the applied standpoint, 1 m$^2$ RCS for the X-band is considered a good objective. For comparison, a small UAV (e.g., DJI Mavic2) has 1000 times smaller RCS, i.e., 10 cm$^2$, which makes it, as is, a very challenging target for surveillance.

Within the realm of advanced scattering control techniques, a special category of highly resonant structures has emerged - "superscatterers".[35–42] As a guideline definition, superscatterers are subwavelength structures ($2R/\lambda < 1$, where $R$ is the minimal radius of a sphere, encompassing the structure and $\lambda$ operational wavelength), that demonstrate scattering cross-sections above the so-called single channel limit, i.e., above the maximum of what a lossless subwavelength resonant dipole can produce ($3\lambda^2/(2\pi)$).[43] Superscatterers typically go against the well-accepted guidelines yet not restrictive criteria, e.g., Chu-Harrington, Geyi, and several others, which suggest a significant operational bandwidth degradation and the scattering cross-section drop with the size reduction.[44,45] The primary operational concept of superscatterers, that allows them to bypass the above-mentioned constraints is resonance cascading. Superimposing several resonant multipoles of the structure allows building high scattering cross-sections, while a considerable 10% fractional bandwidth and above can be achieved by placing the resonances side-by-side.[46,47] The main factor, limiting superscattering performance, is an inherent strong near-field accumulation, which makes those structures extremely susceptible to material losses and intolerant to fabrication imperfections.[36] From the theoretical standpoint, the design of many overlapping and strongly interacting resonances requires performing intensive optimization over multidimensional search spaces. This fundamental requirement calls for exploring advanced algorithms. Among a variety of existent options, genetic algorithms have emerged as one of the most promising methods in the design of superscatterers. By defining the parameters of the structure, specifying degrees of freedom, and formulating a target objective function, genetic algorithms iteratively converge toward acceptable solutions. This approach also enables the revealing of new geometries if topology optimization is undertaken.[48,49]

Here, we investigate 3D highly-scattering architectures based on arrays of near-field coupled electric and magnetic resonators.[50,51] To some degree, those can be considered as tailor-made metamaterials. It will become evident that compared to genetically-designed surfaces, the volumetric dimension has a significant impact on performance.[52]

The manuscript is organized as follows: the design algorithm is presented first, and its outcomes are analyzed. Performances of several realizations are then assessed, and the principle of resonance cascading is revealed. Subsequently, all the theoretical results are verified experimentally, assessing several structures in an anechoic chamber. The applied outlook contains an analysis of a drone, labeled with scattering structures. The numerical predictions are supported with an outdoor experiment, demonstrating the unambiguous advantage of small target tagging with genetically designed metasurfaces and metamaterials. This concept is summarized in Figure 1.

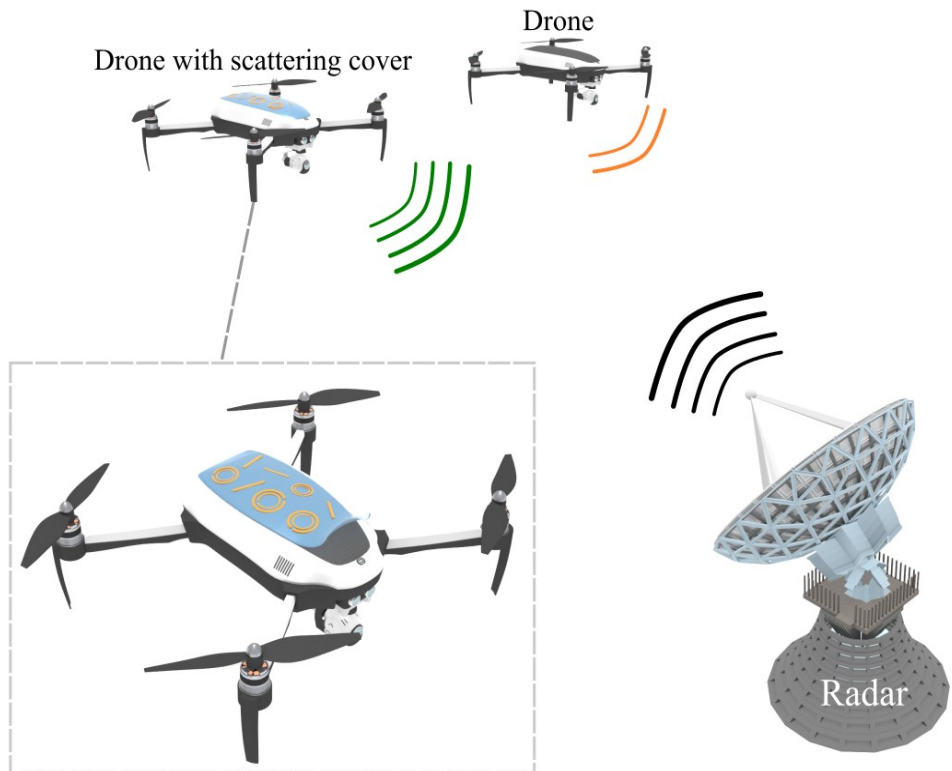

**Figure 1.** The concept of genetically designed surfaces and metamaterials to increase radar visibility of airborne targets – the labeled drone becomes subject to reliable monitoring.

# Genetically designed Metamaterial: Architecture and the Optimization Algorithm

The overall design of the structure consists of stacked-up layers, conforming to the conventional lithographic fabrication process used in making printed circuit boards (PCBs). Layers are divided into pixels, and each one can hold either a wire or a split ring resonator (SRR), essentially functioning as an electric or magnetic dipole. The resonance and spatial orientation of these dipoles can be adjusted by varying their geometry. The spacing between the layers is kept as an additional parameter for optimization. To emphasize, the lateral dimension (referred to as the board side in antenna terminology) of the structure is not subwavelength. On the other hand, the axial dimension (the overall thickness, relevant for end-fire operation) is comparable to a wavelength. This arrangement perfectly complies with the application under consideration, namely radar interrogation of drones, flying at low altitudes (Figure 1 for the concept).

The search space for optimization is as follows: each pixel can encompass either a wire or ring (a discrete variable), the resonator can move in respect to the pixel center (2 continuous variables), the resonator can obtain a rotation in the plane on respect to the global coordinate system (1 continuous variable). Overall, each pixel encompasses 1 discrete and 3 continuous degrees of freedom. Given a 3x3 layer, 3 layers stack-up, the search space becomes 135-dimensional, motivating to consider an efficient optimization routine. The distance between the layers is 10 mm fixed.

The objective function has been tailored to align with the standard requirements of a typical monitoring system, i.e., maximizing radar scattering cross-section (RCS) over a considerable frequency range. Typically, 10% fractions bandwidth complies with the requirements. Here we focus on the X-band (8-12 GHz), which is widely used for radar surveillance and monitoring, due to its effective balance between range and resolution.[53]

Our study uses the Covariance Matrix Adaptation Evolution Strategy (CMA-ES) evolutionary optimization algorithm to obtain the structure with the maximum backscattering.[54] The objective function was defined as the average scattering across N frequencies. These frequencies were recognized by applying two methods: random searches in a specific frequency range and uniform distribution of frequency points in a range 8-12 GHz. Such an objective function allows all frequencies to contribute equally to the final value, enhancing the likelihood of resulting broadband structures without any frequency dominating.

Due to the large number of optimized parameters, additional code optimizations were required to conduct more independent experiments and statistical analysis. To enhance optimization efficiency by performing more iterations in a shorter timeframe and thereby achieving faster convergence, we have incorporated distributed computing over the original evolutionary algorithm. The main resource-demanding operation is the forward solver, which performs the backward RCS calculation. We utilized a customized method of moments approach (PyNEC, based on NEC code), particularly suited for structures composed of conducting wires. In this scenario, the scattering problem is considerably simplified by focusing on determining current distributions on curved wires. This approach is more efficient than full-wave finite elements or finite-difference time-domain methods, typically found in commercial software. Note, that, in the future, for adopting the direct optimization method on a radar target, Green's functions have to be calculated numerically.

To increase the number of experiments and enhance our algorithm's performance, we focused on accelerating the calculation of the objective function, which is the most resource-intensive operation in the optimization process. The increase in structural elements (here, it is an increase in our structure's layers or order) dramatically impacts calculation speed. Given this, we chose to distribute calculations within each of the evolutionary algorithm's generations. This enables parallel calculation of objective functions for several individuals.[55]

Figure 2(a) presents a scheme of a unit cell used during the optimization of the structure. The unit cell is a square area of size 20 x 20 mm. Initially, one of the two possible elements is placed in the center of each unit cell: a double split ring or an electric dipole. The outer radius of the split ring can vary from 5 to 8 mm. The thickness of the ring's conductor is 0.5 mm, the gap between the inner and outer rings is 0.5 mm, and the gap in both rings is 1 mm. The length of the electric dipole can vary from 9 to 18 mm, with a conductor thickness of 0.5 mm. Each element can shift from the center of its unit cell in the plane within a radius δ, which is set before the start of optimization. A value of δ = 0 corresponds to elements tied to the centers of their cells, and δ = 1 means that each element can shift by half its size (t / 2). However, it should be noted that in this case, electrical contact, or interception between elements from different cells is possible. Such cases lead to the termination of the optimization, so a value of δ ~ 0.8 was found to be optimal for the calculations.

An example of the algorithm's implementation of a three-layer structure with 3x3x3 elements after optimization is presented in Figure 2(b) (top and bottom layers are presented). Figure 2(c) illustrates the key steps of a genetic algorithm, specifically mutation and crossover.

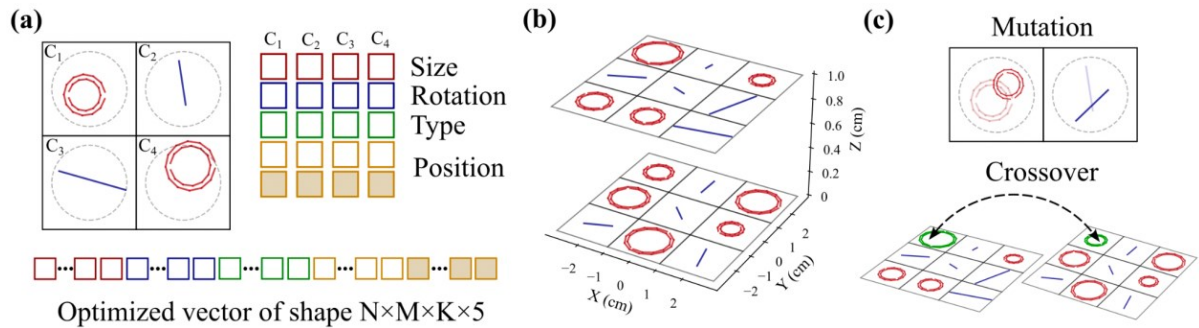

**Figure 2**. Genetically designed metamaterial - optimization algorithm. (a) The search space encompasses continuous and discrete variables. 135 independent degrees of freedom are involved in 3x3x3 structure design. (b) Cartoon of the metamaterial layout – only the upper and lower planes are shown for the sake of clarity. (c) Key steps of the genetic algorithm, specifically mutation and crossover.

The result of the algorithm is the set of designs. We considered 2x2x2 and 3x3x3 structures. Designing larger-scale geometries demands an overuse of computational resources. The times and number of iterations, required for convergence, will be discussed next alongside the discussion on electromagnetic performances.

## Assessment of the Optimized Structures

Structures with different numbers of elements have been designed. The optimization of a 2x2x2 structure takes about one hour, while a 3x3x3 structure requires approximately five hours on a 64-core computational server. This runtime indicates both the computational complexity and guidelines on whether considering larger structures is worth the effort.

Figure 3(a) presents the optimization results for 2x2x2 as a function of the structure's volume, normalized to the cube of the resonant wavelength. The volume is directly related to the unit cell size. The side of a unit cell '$t$' was varied in the range from 60 down to 25 mm. Its further size reduction is challenging as the resonators can start intersecting. The distance between the layers remained fixed at h = 10 mm to keep the end fire cross sections of the structure below the wavelength in width. The displacement value of the elements was $\delta \sim 0.8$. All optimizations were performed for the central frequency of 10 GHz ($\lambda$= 30 mm). For each volume, the maximum number of iterations was fixed as 1000 and conducted independently for 10 different initial seeds. "Error" bars in Figure 3(a) show the spread of the realized objective function (RCS) for the initial seeds. This dispersion highlights the nature of the genetic algorithm, which is not guaranteed to converge to a global extremum. Justification on an optimal number of seeds will follow.

Similarly to the above-described process, 3x3x3 structures were optimized. Here, the side of a unit cell '$t$' was varied in the range from 25 to 95 mm (Figure 3(b)). Comparing the results allows us to conclude that RCS increases with its volume up to a certain point, after which the interaction between elements becomes weaker, and further increasing the size of the unit cell only worsens the outcome. In terms of numbers, 1 $m^2$ RCS is obtained in relevance to the initial motivation of the study.

To verify the results, the structures with the best performance (indicated by red circles in Figure 3(a and b)) were analyzed with CST Microwave Studio (Time Domain solver (FIT method)). 'Best' refers to the highest RCS for 2x2x2 and a compromise between RCS and volume for 3x3x3. RCS spectra of the best structures appear in Figure 3(c and d). CST results were assessed versus the pyNEC code, used as the forward solver, verifying its reliability. Single channel and superradiant limits (single channel limit time the number of the resonators) are indicated as black and blue dashed lines on the plots.[56]

The superradiant limit is overcome by more than an order of magnitude, revealing the constructive near-field coupling between the resonators. Insets in the Figure 3(c and d) demonstrate the actual arrangements within the structure. Endfire incidence is considered and both electric and magnetic field polarization are set to interact with wires and rings, respectively. Worth noting the considerable bandwidth of the structures, reaching 13% and 10% at half-width-half-maximum is achieved and complies with the standards of most surveillance radars, that are in use nowadays.

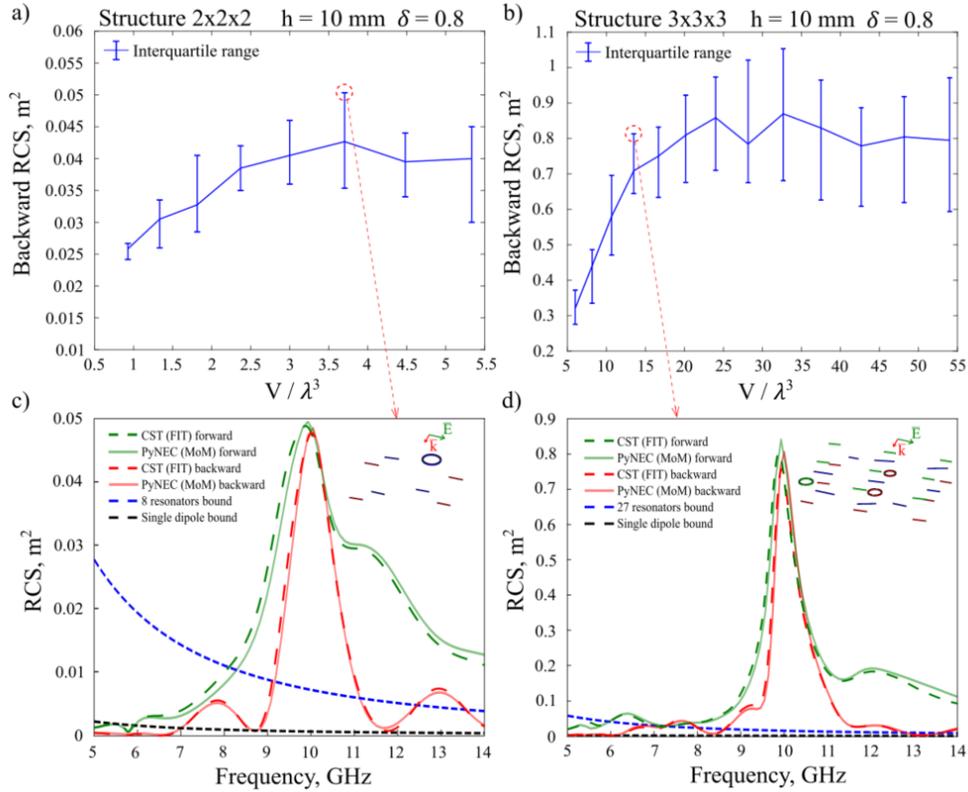

**Figure 3.** RCS optimization for 2x2x2 and 3x3x3 structures. (a), (b) RCS as the function of the normalized structure's volume. 'Error' bars are the results for different (10 in overall) seeds, i.e., seed's dispersion. (c), (d) RCS spectra, calculated with CST and pyNEC, details in legends. Black and blue dashed lines are the single channel and superradiant scatting limits, respectively. Insets - the actual arrangements within the structure, endfire incidence is considered and both electric and magnetic field polarization are set to interact with wires and rings, respectively.

## Assessment of the Algorithm Parameters

From an algorithmic perspective, investigating the impact of the number of seeds on performance is valuable. Figure 4(a) illustrates the RCS span as it varies with the number of seeds. The results indicate that 10 seeds suffice for reasonable optimizations, while 30 seeds are enough to approach close to the maximum performance. In the analyses above, 10 seeds were used as a compromise between the computational time and performance. Figure 4(b) reveals the structure of the seed dispersion. 100 seeds were chosen, and an RCS histogram was plotted. To an extent, the histogram resembles a Gaussian distribution. Worth saying that this is not trivial as central limit theorem conditions are not applicable here.

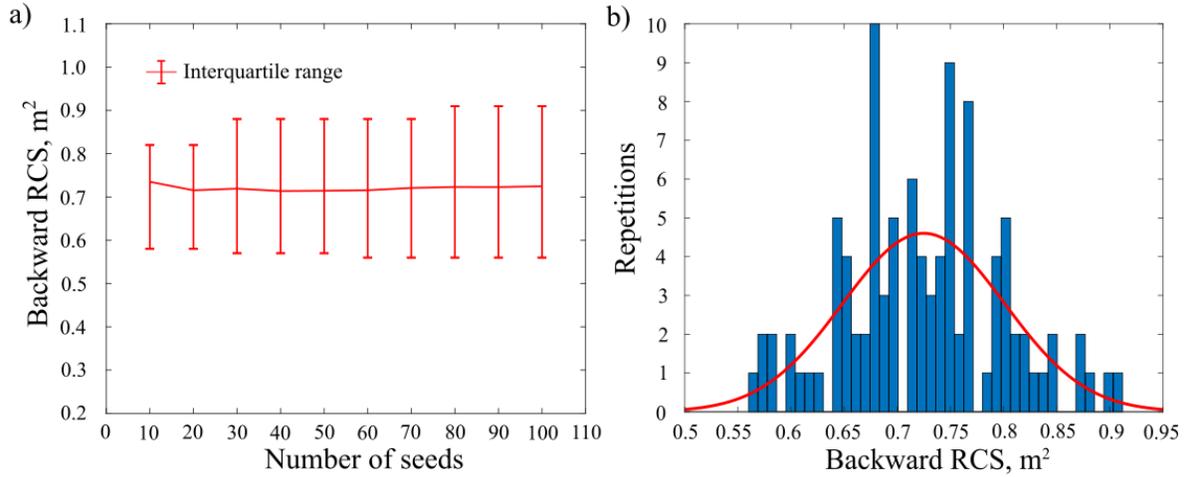

**Figure 4.** (a) RCS span as it varies with the number of seeds. (b) Histogram, highlighting the internal structure of the seed dispersion (for 100 seeds). Red curve - Gaussian fit.

## Experimental Results

The structures were fabricated with a conventional printed circuit board (PCB) methodology. For clarity, the schematics of a typical layout, including the substrate appear in Figure 5(a). To minimize the effect of a substrate's permittivity on the resonant response, the sample was fabricated by chemically etching copper elements on a very thin dielectric board: Rogers 3003 ($\varepsilon_r = 3$, $\tan(\delta) = 0.001$) with a thickness of 0.127 mm has been used. The photograph of the sample is presented in Figure 5(b). Styrofoam, transparent for GHz waves, is used as a spacer between the PCBs.

The measurements were taken in an anechoic chamber. The experimental setup included a broadband horn antenna NATO IDPH-2018 (serving for both Tx and Rx – transmit and receive, certified for 2-18 GHz frequency range), which was connected to the Keysight P9374A Network Analyzer 300 kHz - 20 GHz (photograph of the scheme is presented in Figure 5(c)). For obtaining quantitative measurements, a 10 cm brass calibration disk has been used. Time gating post-processing was applied to further reduce the multipath impact. Figures 5(d) demonstrate both numerical and experimental backscattering cross-sections (in $m^2$) spectra, demonstrating the distinguished peak around 10 GHz, for which the structure was originally optimized. 0.6 $m^2$ RCS is demonstrated in the experiment for the end-fire excitation of the device. Given the multiple overlapping resonances and the structure's high sensitivity to fabrication tolerances and substrates, the agreement between the experimental and numerical data is considered satisfactory. To recap, this number is ~3 orders of magnitudes larger than the RCS of a small drone.

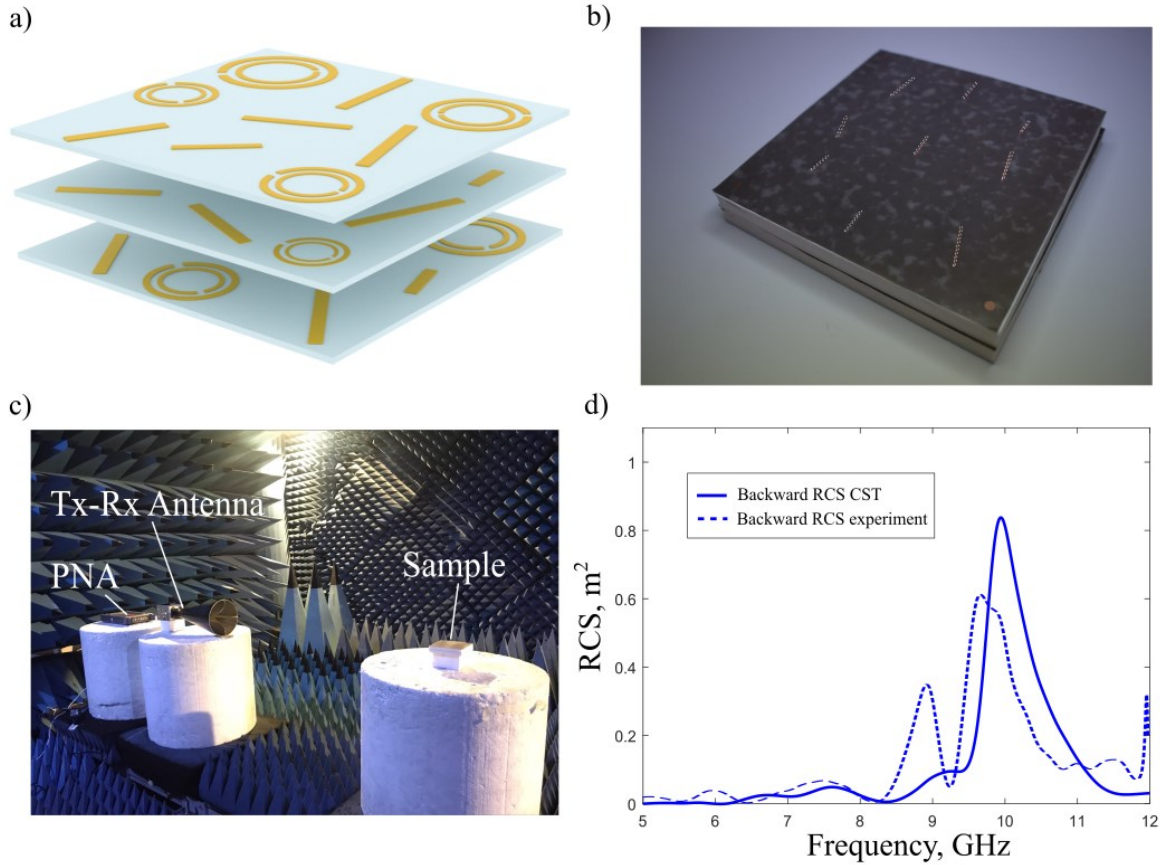

**Figure 5.** Characterization of genetic metamaterial RCS. (a) schematic layout of 3x3x3 structure. (b) Photograph of the experimental sample. (c) Photograph of the experimental sample – backscattering measurements in an anechoic chamber. (d) Experimental and numerical RCS spectra.

## Drone Labeling, Numerical Assessment

After demonstrating the fundamental scalability of genetic metamaterials in scattering management, an application will be considered. To enhance the inherently low RCS of the DJI Mini 2 drone used in tests, the UAV is equipped with the scattering structure. Considering low flight altitudes (~50 meters) and the location of a surveillance radar (several kilometer distances), the end-fire operation of the device is well justified (the observation angle is on a degree scale). The full-wave electromagnetic modeling with the STL model of the DJI Mini 2 drone was performed in CST. As for the model, the drone was considered as a solid, made of plastic with permittivity $\varepsilon = 3$ and loss tangent $\tan(\delta) = 0.02$, corresponding to tabulated values.[57]

Two interaction scenarios were assessed, for both the elevation angle is 0°, complying with the surveillance scenario (the k-vector of the incident wave and the horizontally oriented drone belong to the same plane). In-plane angles of incidence are 0° and 45° while the wave is the z-axis (see Figure 6(a)). Four scenarios were considered, and the scattering spectra appear in Figure 6, details are in legends. Several observations can be deduced. First, the RCS of the tagged drone is dramatically higher compared to the untagged counterpart (by a factor of 52.7). Considering the radar equation, the detection range which scales with the 4th root to the scattering cross-section, grows by 2.6 times.[58] The peak of the enhancement at 9.4-10GHz coincides with the structure's resonance. Furthermore, it is quite remarkable that significantly lower RCS is obtained for 45° angles of incidence. From the physical point-of-view, it is related to the strong angular sensitivity of the scattering structure. This effect also has an important applied significance. In the realm of multiple moving targets, radar prioritizes only those which move towards it. In this scenario, RCS enhancement is only valuable for approaching targets, while low-priority UAVs do not provide false alarms and introduce noise into the detection channel. In this case, the RCS enhancement owing to tagging is only by a factor of 2. The strongly directive scattering pattern can be observed from the analyses, which appears in onsets in Figure 6(b). This aspect also influences scenarios where multiple radars, forming a radar chain, are used for monitoring. In such setups, directive patterns ensure that there is no interference with neighboring radars.

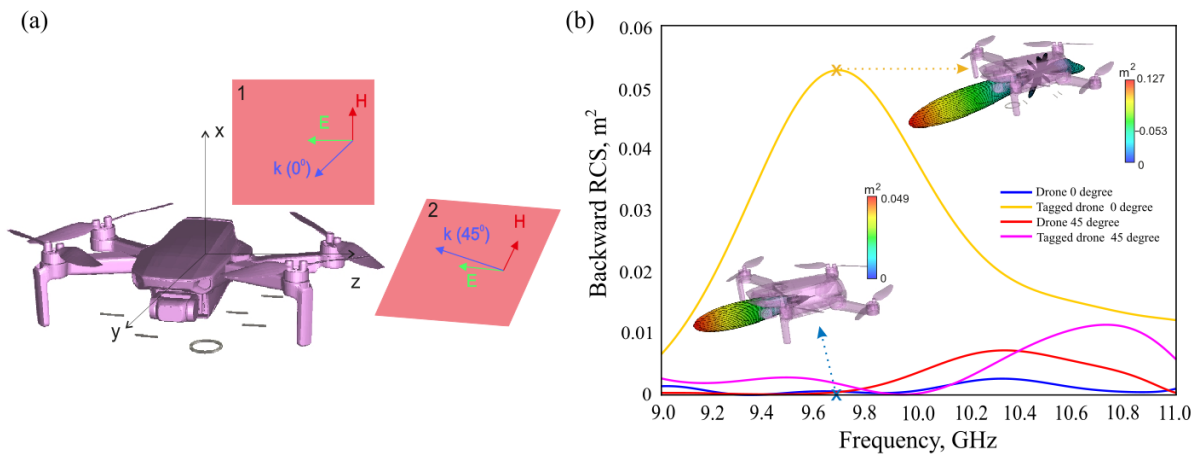

**Figure 6.** (a) The interaction scenario with DJI Mini 2 drone. (b) RCS spectra of the drone. Details are in the legend. Insets show the scattering diagrams.

To explore the operation principle of the structure, the multipole expansion of the scattering cross-section is conducted. Specifically, in this context, the resonance cascading principle can be revealed. The structures were analyzed based on the routine, reported recently.[59] Spherical multipoles up to 8[th] order were calculated using surface integral formulation. Figure 7(a) illustrates the multipole expansion, with the black line representing the sum of all multipole fields of the system (total scattering), a representation consistent with results obtained with CST and Comsol. The different colors indicate the contributions of multipole fields arranged within the frequency range of interest. The primary contributions are made by the electric dipole (ED) and magnetic quadrupole (MQ) components.

To reveal the impact of the drone body on scattering, additional electromagnetic modeling was conducted using an STL model of the DJI Mini 2 drone both with and without our 3x3 structure. Figure 7(b) shows the electric field maps in the XY plane, cutting through the middle of the structure. A plane wave was directed along the y-axis and polarized along the z-axis. It can be seen that when only the drone is present, the plane weakly interacts with the drone body owing to the relatively low refractive index contrast. However, equipping the UAV with the 3x3x3 scatterer is dramatically boosting the interaction, resulting in a very strong scattering.

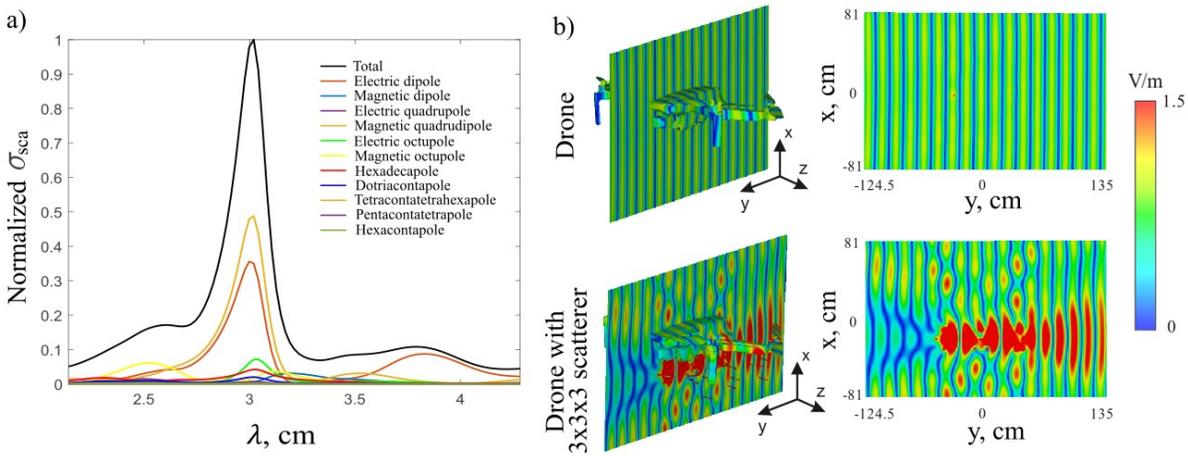

**Figure 7.** (a) Multipole expansion of 3x3x3 scatterer. The spherical multipolar contributions are in legends. (b) Total electric field intensity in XY-plane cutting through the UAV's center. The top and bottom plots correspond to the standalone drone and the one equipped with the scatterer. The fields are in scale.

## Labeled Drone Scattering, In-door Experiments

Experimental tests were performed by attaching the scattering structures (both 2x2x2 and 3x3x3) to the bottom of the DJI Mavic 2 and comparing the results with the untagged drone. The experimental layout appears in Figure 8(a), where the monostatic scheme is explored considering the 5-12 GHz frequency range. Figure 8(b) summarizes the results. First, as was previously predicted, the labeling dramatically increases the structure's RCS. The main enhancement peak is close to 10 GHz, as it was initially designed.

It also encompasses more than 10% fractional bandwidth. The 3-fold difference between 2x2x2 and 3x3x3 is not very dramatic, thus the smaller structure can be favored owing to its better aerodynamic properties.

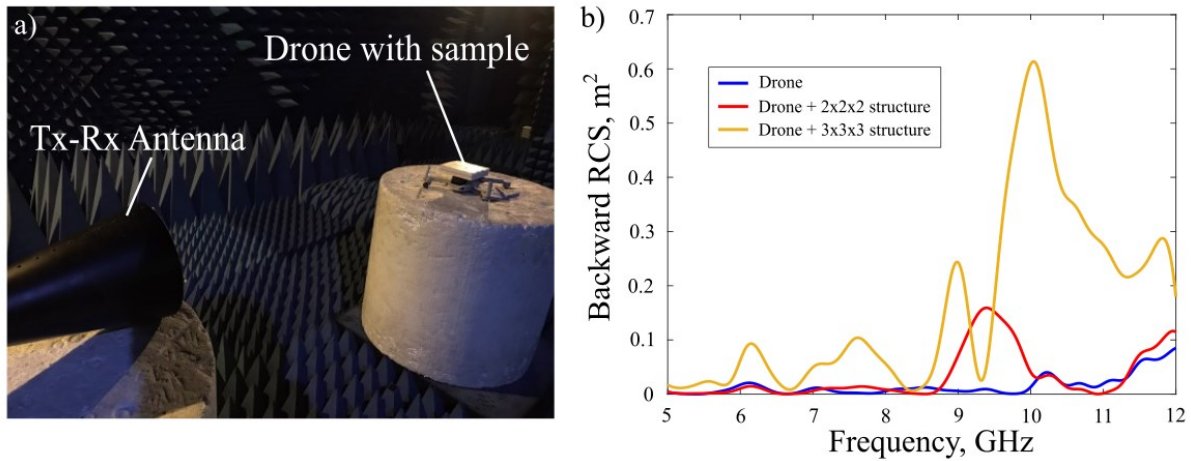

**Figure 8.** (a) Photo of experimental measurements of backward scattering from DJI Mini 2 drone (blue line) with 2x2x2 structure (red line) and with 3x3x3 structure (orange line).

## Labeled Drone Scattering, Outdoor Monitoring

Considering the form factor of the UAV and the payload capacity, it can carry only the 2x2x2 scatterer, which was assessed experimentally. Outdoor experiments (Figure 9(a)) were performed with a custom-made continuous wave (CW) radar, based on Keysight P9374A Network Analyzer 300 kHz - 20 GHz (Figure 9 (b)). The 2-ports PNA, operated by Matlab, was connected to Tx and Rx broadband horn antennas (NATO IDPH-2018, 2-18 GHz). Horizontal polarization relative to the terrain was maintained. The output power (at Tx) was 20 dBm. A low-noise amplifier (LNA, MWA-020180-1-4019, 2-18 GHz) was applied at the Rx. To avoid a feedback loop between transmit and receive channels, the antenna isolation was maintained at more than a 30 dB level, which was checked by monitoring the $S_{21}$ spectrum before introducing the drone. This is the reason why Tx and Rx were taken apart, compared to the laboratory experiments. The bandpass filter (VHF-6010+, 6.3-15 GHz) was used to suppress noise at frequencies irrelevant to the experiment. Photographs of untagged and tagged drones appear in Figure 9 (c and d).

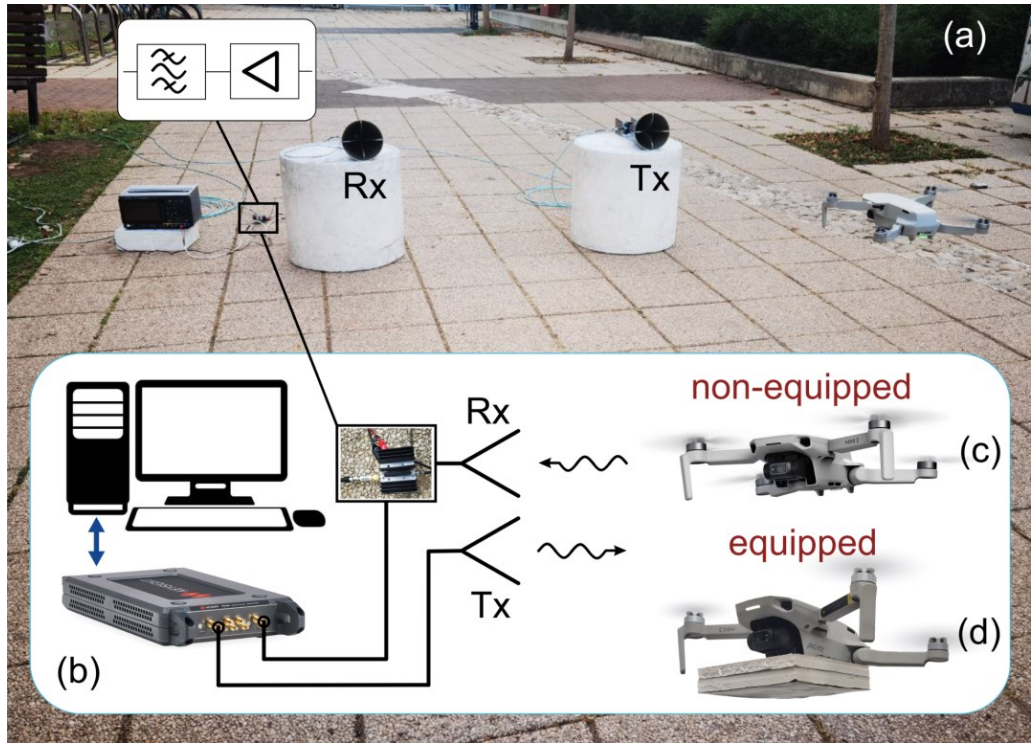

**Figure 9.** (a) Experimental setup for outdoor detection of drones using a custom-made CW (Doppler) radar. (b) Schematic of the custom-made radar. Photos: (c) untagged drone and (d) drone equipped with 2x2x2-structure.

To assess the impact of the resonant structure on the drone visibility and detectability, the central frequency of the CW radar was swept in the range between 7 and 11 GHz – in this region, the RCS changes for the most (Figure 9(b)). The carrier frequency was swept with a 200 MHz step. During each test flight, the radar recorded complex-valued (real and imaginary) signals for 10 seconds, sampling the range at 60001 points. The IF (intermediate frequency) bandwidth was set to 30 kHz. To improve the robustness, each flight was repeated 3 times. Before each flight, the drone was positioned 40 meters in front of the radar, which was triggered by launching. The radar failed to detect the UAV after it passed by and exited the antenna beam.

The baseband signals are presented with spectrograms, which appear in Figure 10. A manually adjusted bandpass filter was applied to remove slow-moving clutter and fluctuations at higher frequencies. The spectrogram settings are 75% overlap between adjoining segments, a time window of 0.33 s, and a 10-second observation time on the moving target. The visual analysis of the spectrograms clearly shows that in the resonant case (9.6 GHz, Figure 8(b) the tagged done is seen for at least another 2 seconds of flight (comparison between panels a and b). The motion starts at $t=0$ (corresponding to 40 m distance from the radar) and the drone is not visible, i.e., there are no pronounced frequencies on the map. 2 seconds of flight correspond to ~doubling in the detection distance, which, following the radar equation, is related to x16 enhancement in RCS. This number is well-related to the experimental measurements, presented in Figure 8(b). Note, that the scattering from the tagged drone is highly directive. Thus, the drone, flying at an altitude above the radar (Figure 9 (a)), becomes less visible when approaching the antenna. At large distances, the interaction is end-fire, while at a closer proximity, the angles start to contribute. This is the reason, why the Doppler map of the untagged drone is brighter at shorter distances. At $t = 7.5$ s the drone passes above the antenna and disappears from the detection. This is seen for times $t > 8$ s.

For 10.8 GHz, both tagged and untagged drones have similar RCSs, thus the detection distances are similar to each other and also significantly shorter for the resonance frequency. Worth noting, however, that low-altitude flights are subject to significant clutter and multipath aspects. In this case, quantitative comparison between performances at different frequencies becomes more challenging (this point will be addressed hereinafter). Flying at high altitudes, i.e., open sky conditions, requires different apparatus and is also prohibited on campus.

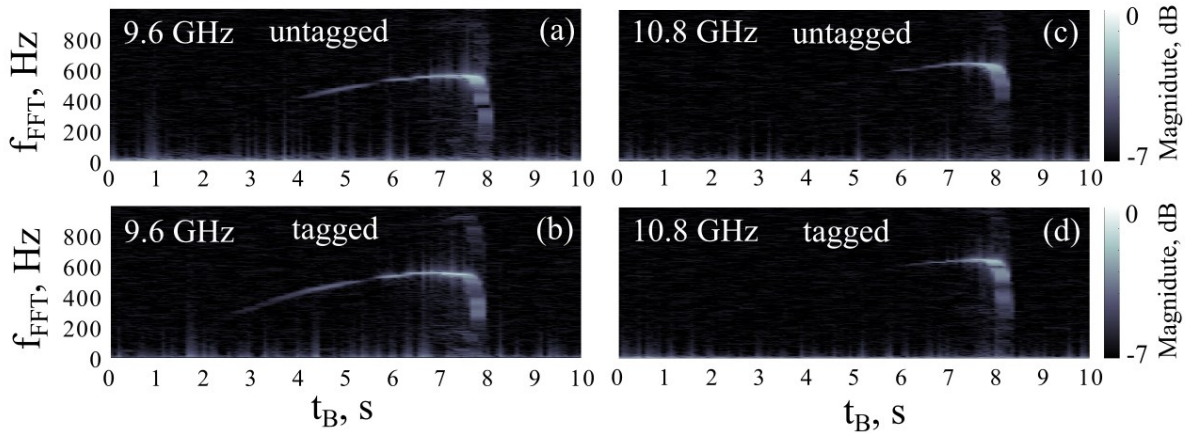

**Figure 10.** Short-Time Frequency Transform for the untagged and tagged flights at different carrier frequencies of the used CW radar. 9.6 GHz corresponds to the regime, where the 2x2x2 tag has the maximal RCS. For 10.8 GHz tagged and untagged drones have a similar RCS.

Finally, to explore additional capabilities of tagging and relate it to real distances, the following analysis has been done. First, the measured drone velocity (e.g., from Doppler maps in Figure 10) was translated to the distance in meters considering the nonuniform accelerated motion. This mapping with the CW radar is made possible since the target flies toward the radar and co-locates with it at a known time. The Doppler filter was adjusted to fit the 7-9m/sec drone's velocity (a priori known information in the experiment) and +3dB SNR was chosen as a threshold.

The vertical cut at noise 0 dB level (the left side of the map) on the colormap in Figure 11(a) shows the maximal detectability distance of the untagged drone as a function of carrier frequency (vertical axis). To assess the impact of the SNR degradation, white noise was manually (digitally) added to the raw data from the measurements. The horizontal axis on the colormap is related to the additive noise level. 0 dB is no noise added (the right part of the plot), and 70 dB noise (on the left side) is the loser limit. For example, adding such excessive noise does not allow drone detection – blue areas on the color map. Figure 11(a) is the detectability map for the untagged drone, while panel (b) is the data for the tagged sample. Figure 11(c) is the difference between panels (b) and (a). It can be clearly seen that tagging significantly enhances the detectability range for carrier frequencies, where the scatterer has resonance. The secondary hot spot in Figure 11(c) appears around 10.5 GHz and, most likely, is a multipath and interference artifact, related to low-altitude flights.

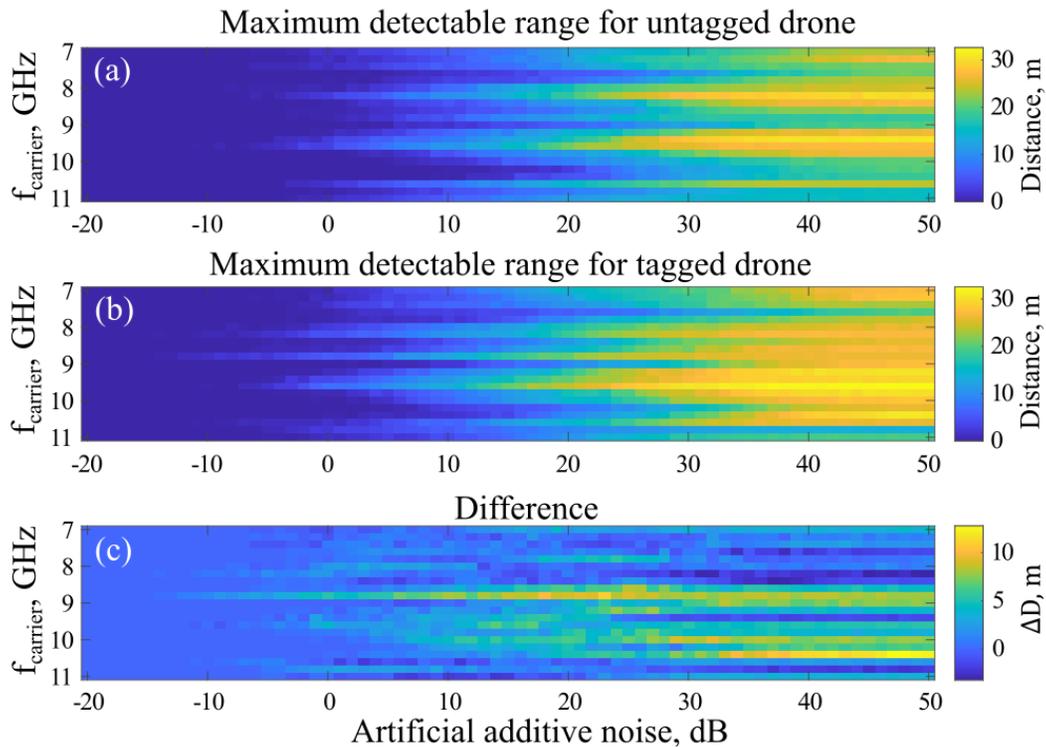

**Figure 11.** Colormaps of the maximal detectability range as the function of the carrier frequency (vertical axis) and the artificially added noise (horizontal axis). 0dB stands for the raw data, collected during the outdoor measurement. (a) Untagged drone. (b) Tagged Drone. (c) difference between (b) and (a).

## Conclusions

We presented a novel concept of genetically engineered metamaterials, tailored to maximize scattering in end-fire incidence scenarios. These structures feature multi-layered arrays containing near-field coupled electric and magnetic resonators. Multi-objective genetic optimization to boost the radar scattering cross-section at relatively high fractional bandwidth has been implemented and provided designs of structures featuring a broadband scattering of ~ 1m² at 10 GHz with over 10% fractional bandwidth. Alongside its compact broadside dimensions, metamaterial scatterers exhibit a small end-fire physical cross-section which is below a single squared wavelength. Among a variety of different practical applications, UAV monitoring in urban environments was chosen as for the demonstration. Specifically, we have demonstrated that initially small 10 $cm^2$ RCS of a small drone can be boosted by 3 orders of magnitude by equipping it with the metamaterial scatterer. Outdoor experiments demonstrated a two-fold detectability range extension with a very small metamaterial scatterer, attached to the drone body. Lightweight, conformal accessories with significantly high scattering cross-sections, are designed to enhance the capabilities of passive monitoring systems, and can add a vital layer of security in urban airspace. These add-ons can be specifically engineered to support surveillance infrastructure by improving detectability and coverage.

## Acknowledgments

Department of the Navy, Office of Naval Research Global, under ONRG Award N62909–21–1–2038, Israel Science Foundation (ISF grant number 1115/23), and Israel Innovation, Authority.